\documentclass[10pt, conference]{IEEEtran}

\let\labelindent\relax

\ifCLASSOPTIONcompsoc
  \usepackage[nocompress]{cite}
\else
  \usepackage{cite}
\fi
\ifCLASSINFOpdf
  \usepackage[pdftex]{graphicx}
  \graphicspath{{./images/}}
  \DeclareGraphicsExtensions{.jpg}
\else
 \usepackage[dvips]{graphicx}
\fi
\usepackage{amsmath}
\usepackage{amssymb}

\usepackage{url}

\usepackage{lipsum}
\usepackage{dirtytalk}
\usepackage{subcaption}
\usepackage{amsfonts}
\usepackage{color}

\usepackage[inline]{enumitem}

\usepackage{tikz}
\usetikzlibrary{calc,positioning,arrows}

\newcommand{\imageref}[4][0.5]{
  \begin{figure}
    \includegraphics[width=#1\textwidth]{#2}
    \caption{#3}
    \label{fig:#4}
  \end{figure}
}

\usepackage{xcolor}
\usepackage{listings}

\newcommand\JSONnumbervaluestyle{\color{red}}
\newcommand\JSONstringvaluestyle{\color{red}}

\newif\ifcolonfoundonthisline

\makeatletter

\lstdefinestyle{json}
{
  showstringspaces    = false,
  keywords            = {false,true},
  alsoletter          = 0123456789.,
  morestring          = [s]{"}{"},
  stringstyle         = \ifcolonfoundonthisline\JSONstringvaluestyle\fi,
  MoreSelectCharTable =%
    \lst@DefSaveDef{`:}\colon@json{\processColon@json},
  basicstyle          = \ttfamily,
  keywordstyle        = \ttfamily\bfseries,
}

\newcommand\processColon@json{%
  \colon@json%
  \ifnum\lst@mode=\lst@Pmode%
    \global\colonfoundonthislinetrue%
  \fi
}

\lst@AddToHook{Output}{%
  \ifcolonfoundonthisline%
    \ifnum\lst@mode=\lst@Pmode%
      \def\lst@thestyle{\JSONnumbervaluestyle}%
    \fi
  \fi
  \lsthk@DetectKeywords%
}

\lst@AddToHook{EOL}%
  {\global\colonfoundonthislinefalse}

\begin{document}
%
\title{Monitoring dynamic mobile ad-hoc networks: A fully Distributed Hybrid Architecture}

\author{

  \IEEEauthorblockN{Jose Alvarez and Stephane Maag}
  \IEEEauthorblockA{
    SAMOVAR, Telecom SudParis, Universit\'e Paris-Saclay\\
    9 Rue Charles Fourier, 91000, Evry, FR\\
    \{Jose\_Alfredo.Alvarez\_Aldana,\\
    Stephane.Maag\}@telecom-sudparis.eu
  }

\and

  \IEEEauthorblockN{Fatiha Za\"{i}di}
  \IEEEauthorblockA{
    LRI-CNRS, Universit\'{e} Paris Sud, Universit\'e Paris-Saclay\\
    15 Rue Georges Clemenceau, 91400, Orsay, FR\\
    Fatiha.Zaidi@lri.fr
  }
}

\maketitle

\begin{abstract}

The mobile ad-hoc networks (MANETs) represent a broad area of study and market interest. They provide a wide set of applications in multiple domains. In that context, the functional and non-functional monitoring of these networks is crucial. For that purpose, monitoring techniques have been deeply studied in wired networks using gossip-based or hierarchical-based approaches. However, when applied to a MANET, several problematics arise mainly due to the absence of a centralized administration, the inherent MANETs constraints and the nodes mobility. In this paper, we present a hybrid distributed monitoring architecture for mobile ad-hoc networks in context of mobility pattern. We get inspired of gossip-based and hierarchical-based algorithms for query dissemination and data aggregation. We define gossip-based mechanisms that help our virtual hierarchical topology to complete the data aggregation, and then ensure the stability and robustness of our approach in dynamic environments. Further, we propose a fully distributed monitoring protocol that ease the nodes communications. We evaluate our approach through a simulated testbed by using NS3 and Docker, and illustrate the efficiency of our mechanisms.

\end{abstract}


%
\IEEEpeerreviewmaketitle

\section{Introduction} \label{introduction}

Wireless mobile ad hoc networks (MANETs) are collections
of mobile nodes that communicate with each other, representing thus a broad area of study and interest.
They provide wide sets of applications in multiple contexts like for instance disaster recovery, military coalition, vehicular ad-hoc networks, sensor networks, and many others.
Besides, network monitoring have been deeply studied in P2P, DTN or decentralized static networks using gossip-based or hierarchical-based approaches \cite{wuhib2009robust,kempe2003gossip,guerrieri2010distributed,artigas2006deca}.
However, when it is applied to a MANET, new problematics arise mainly due to the absence of a centralized administration, the inherent MANETs properties and the node mobility.
Some approaches propose a coordinator, nevertheless, due to energy efficiency, Internet access, infrastructure or various other parameters, these solutions are not always applicable.

To monitor a network, it is often needed to be able to find out the status of the network or some of its properties. 
Generally, an intuitive approach is to define a central node as a coordinator for storage and processing of the observations.
This is notably proposed by \cite{cormode2013continuous}, where the author surveys the different communication mechanisms for an optimal data interchange.
There are multiple studies referring to the centralized approaches that focus on different algorithms to enhance the overall process.
These centralized architectures are efficient for certain types of topologies, but become critical when considering dynamic topologies.
This is why there has been a lot of efforts on decentralized monitoring.

As stated in \cite{stingl2012benchmarking}, the non-functional requirements of a decentralized monitoring mechanism are: performance, costs, fairness, scalability, robustness and stability.
When we refer to performance, we refer to the accuracy of the delivered results and how fast are the results delivered.
The costs refer to the overhead by the communication or processing of the data.
Fairness can be analyzed in respect of performance and cost.
The scalability refers to the ability to work in large and dense networks as well as in small networks.
The robustness deals with the behavior of external and unpredictable events.
And the stability is the ability to address random behavior of autonomous nodes.

Based on these requirements, we can observe the existing solutions which rely on gossip-based or hierarchical-based approaches.
Each of them provides a specific set of features and also downsides for the monitoring process, as studied by \cite{stingl2012benchmarking}.
Gossip-based approaches demonstrate their robustness and stability in dynamic scenarios and changing topologies.
Nonetheless, depending on the scalability, the cost and performance can be impacted.
On the other side, hierarchical approaches show an efficient performance, cost and scalability, although the  robustness and stability may decrease in dynamic scenarios.
This shows that the two major categories perform very good under different characteristics, requirements and constraints of a network.
Therefore, we guess that a more prominent algorithm could be derived from these two approaches for wider scenarios.

The main contribution of this paper is the proposal of a hybrid algorithm for decentralized monitoring of mobile ad-hoc networks in dynamic context.
We define an architecture combining gossip-based and hierarchical-based algorithms for query dissemination and data aggregation.
We perform the gossip-based approach to disseminate the query and in the process to build a virtual hierarchical topology (VHT) for a time window.
Once the query is disseminated through all the network, with the support of the VHT, a hierarchical-based aggregation takes place.
To provide robustness and stability, we provide intermediary gossip-based mechanisms 
to complete the aggregation even in dynamic environments.

The second contribution of this paper is the definition of a monitoring protocol that aims at helping a decentralized monitoring process.
We believe that many approaches propose interesting techniques for querying and aggregating the data, but from our knowledge, no ones propose any definition of a protocol easing the nodes communications.
In order to define our monitoring protocol, we rely on the needs based on our hybrid algorithms and through json (RFC 7159\cite{RFC7159}), we characterize the structure.
Our expectation is not just to provide a structure but also a mathematical background for further model checking and testing. Our protocol has been successfully assessed using NS3 and Docker with mobility patterns. 


The remaining of our paper is as it follows. 
In Section \ref{preliminaries}, we discuss the monitoring, the different types of monitoring and how that influences our decision for our algorithm.
Then in Section \ref{methods}, we present the hybrid algorithm.
To provide a clear explanation we use multiple examples to achieve it.
In Section \ref{results}, we present our implementation, with a semi-formal support for our protocol. 
We also evaluate our approach using NS3 and Docker through a configurable emulated testbed. 
We illustrate and discuss the effectiveness of our mechanisms.
Next, in Section \ref{relatedworks}, we present some interesting related works from which we got inspired. 
After this, in Section \ref{futurework} we present some discussions regarding the future works of our research.
And finally, we conclude our paper by presenting our conclusions in Section \ref{conclusions} .


\section{Preliminaries} \label{preliminaries}

Network monitoring is an extensive field of interest.
It can be described as \say{A number of observers making observations and wish to work together to compute a function of the combination of all their observations} (\cite{cormode2013continuous}). 
The observers are here the nodes in a MANET.
The goal is that all network nodes compute a value $f(t)$ in a given instant of time $t$ in a collaborative way.
The function $t\mapsto f(t)$ [$\mathbb{R}^{+*}\rightarrow X$, $X$ being the domain targeted by $f$], for our purposes, is a linear and non-complex function like the average CPU, 
or any other nominal value.
There are actually multiple definitions but it depends on the focus.
We will present the basic terms for the reader to understand the following sections.



\subsection{Types of Monitoring}

The classification of the monitoring process has been studied in \cite{battat2014monitoring}.
For the purposes of this paper, we will consider two major types: centralized and decentralized.
The centralized type of monitoring, as stated by \cite{cormode2013continuous}, is when all the nodes report their observations to a central entity, named centralizer or coordinator.
This entity will process all the observations from all nodes in order to reach a global view of this property of the network.
The decentralized approach, the opposite of the previously defined approach, deals with networks where there is no centralizer entity.
This implies that the network by itself needs to achieve a global view of a property of the network.
The preferred approach is to get the global view of a property of the network, and then disseminate it through the entire network to assure that is available to all the network.
As stated by \cite{stingl2012benchmarking}, the more noticeable 
approaches are currently gossip and hierarchical.

\subsubsection{Gossip-based approaches}

The gossip approaches are based on the gossip or epidemic algorithms.
Gossip algorithms rely on selecting, from a set of reachable nodes, a random or a specific node (depending on the algorithm) to forward the data packet.
There are multiple types of gossip techniques, a more detailed study can be found in \cite{chakchouk2015survey}.
In this publication, the focus is on routing protocols.
But it explains how gossip based protocols implement through different techniques and properties their goals.
Epidemic algorithms try to forward the packet not only to one but to multiple nodes.
And they also are considered the same or a subcategory of the gossip algorithms.
Flooding is the most common and simple algorithm for epidemic algorithms.
Gossip based monitoring algorithms have the advantage of being highly stable and perform better in increasing dynamic networks.
However, 
it may generate lots of traffic and, under certain scenarios, require more time to compute a value.

\subsubsection{Hierarchical-based approaches}

Hierarchical approaches commonly use tree structures 
in which the leaf nodes communicate the values with their parent node.
This is done recursively until it reaches the root node of the hierarchy.
These approaches consider a mechanism of either pulling data or pushing data with their nearby nodes.
To apply a hierarchical algorithm over a network, it is needed to build the topology before being able to monitor.
The advantages of hierarchical based monitoring algorithms are that they provide a fast convergence of the monitored property and produce less traffic.
The disadvantages of these solutions are that they are prone to errors in the event of a crash in the network. This means that it does not perform efficiently in a highly dynamic environment.

\section{Hybrid Monitoring Approach} \label{methods}

The hybrid algorithm architecture herein proposed consists in two network states, the \say{query state} and the \say{aggregate state}.
The election of the start node is out of the scope of this publication and will be analyzed in future works.
The idea is to combine a gossip approach and a hierarchical approach to achieve the monitoring of a property of the network.
The communication between the nodes to achieve the monitoring of the network will be achieved through a package previously defined.
The idea is that a start node will start the monitoring process by propagating a monitoring query in a gossip approach.
The approach chosen will be described as epidemic, given that the idea is to share information in an efficient, fast and simple way.
Each hop, the nodes will exchange information creating a VHT valid only during this monitoring process.
Then based on this topology, the nodes will start aggregating the information by sending their results to the parent node.
Once the aggregation is done and has reached the start node, there will be a global view of the measured property and the VHT will no longer be usable.
If the process starts again, a new VHT will be derived.

\subsection{Detailed Example} \label{example}

\begin{figure*}
  \begin{subfigure}{.25\textwidth}
    \centering
    \includegraphics[width=.8\linewidth]{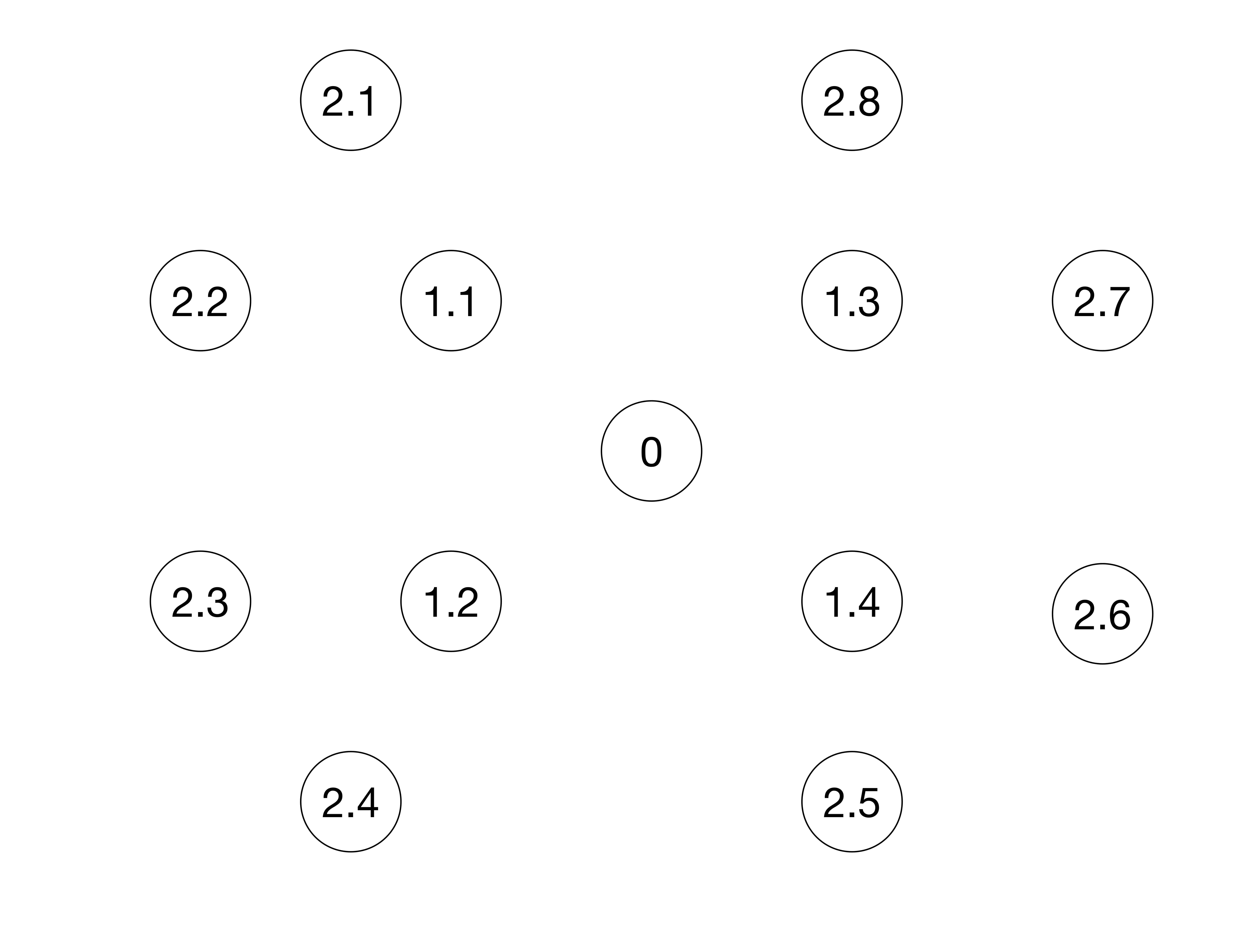}
    \caption{}
    \label{fig:examplesfig1}
  \end{subfigure}%
  \begin{subfigure}{.25\textwidth}
    \centering
    \includegraphics[width=.8\linewidth]{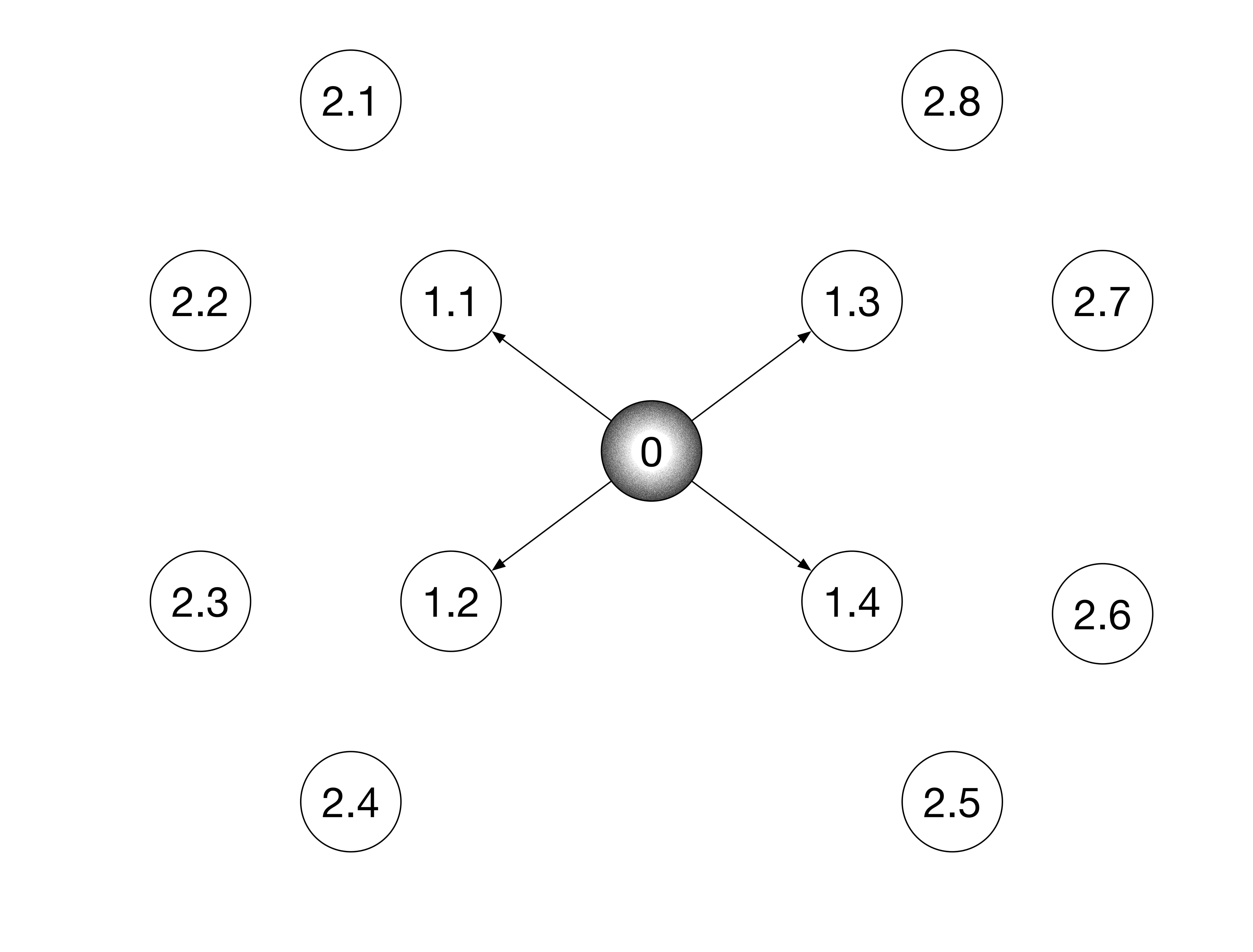}
    \caption{}
    \label{fig:examplesfig2}
  \end{subfigure}%
  \begin{subfigure}{.25\textwidth}
    \centering
    \includegraphics[width=.8\linewidth]{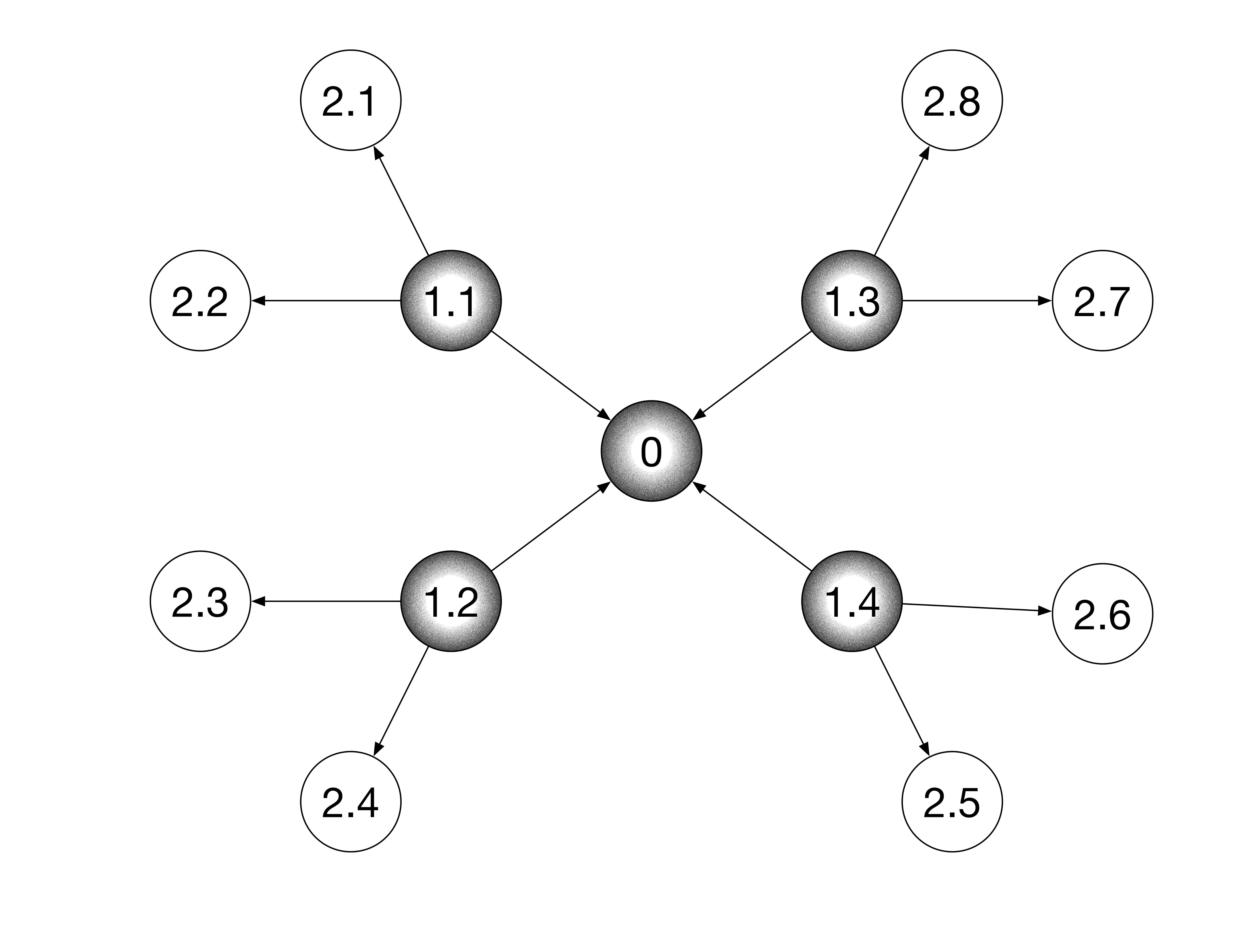}
    \caption{}
    \label{fig:examplesfig3}
  \end{subfigure}%
  \begin{subfigure}{.25\textwidth}
    \centering
    \includegraphics[width=.8\linewidth]{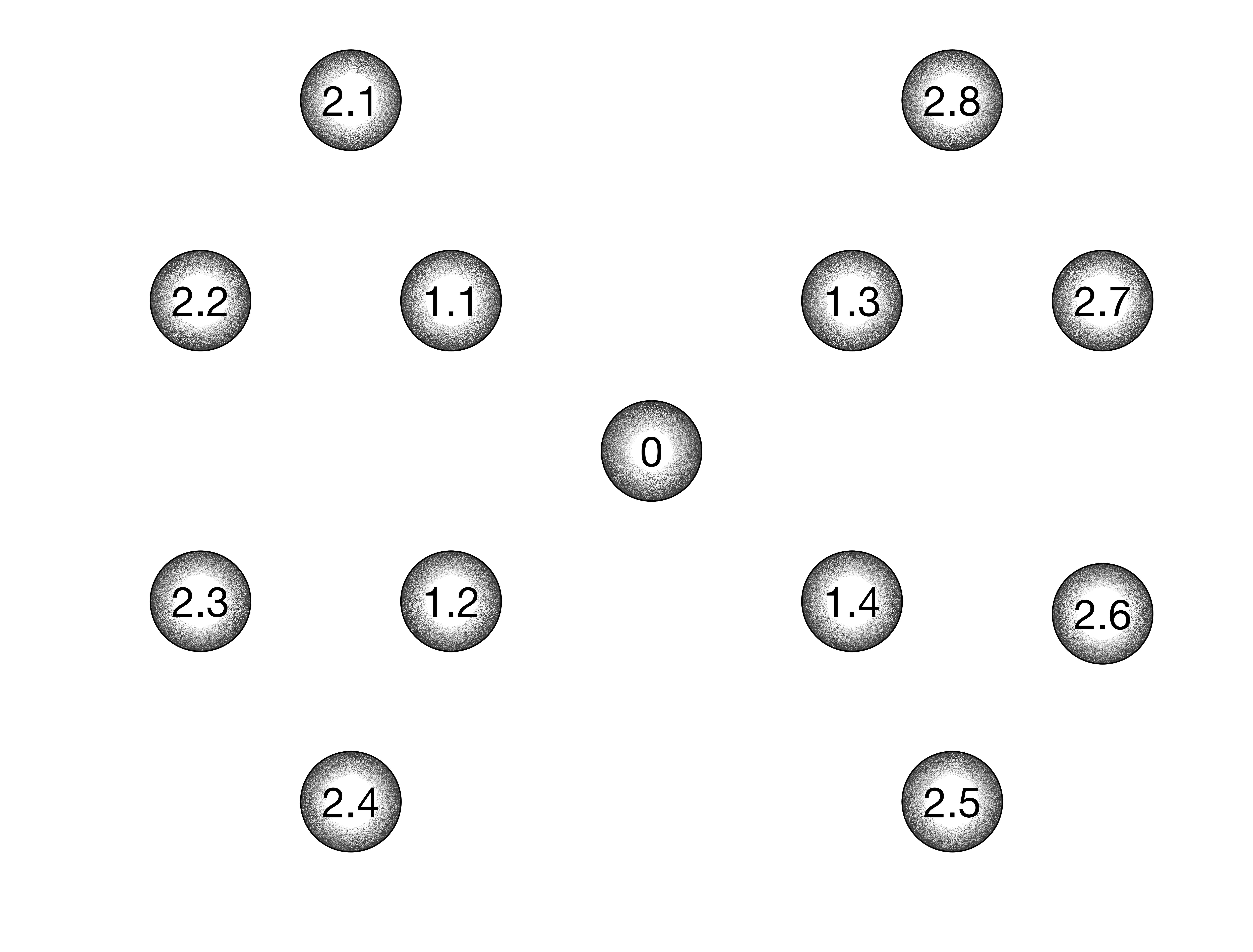}
    \caption{}
    \label{fig:examplesfig4}
  \end{subfigure}
  \par\medskip
  \begin{subfigure}{.25\textwidth}
    \centering
    \includegraphics[width=.8\linewidth]{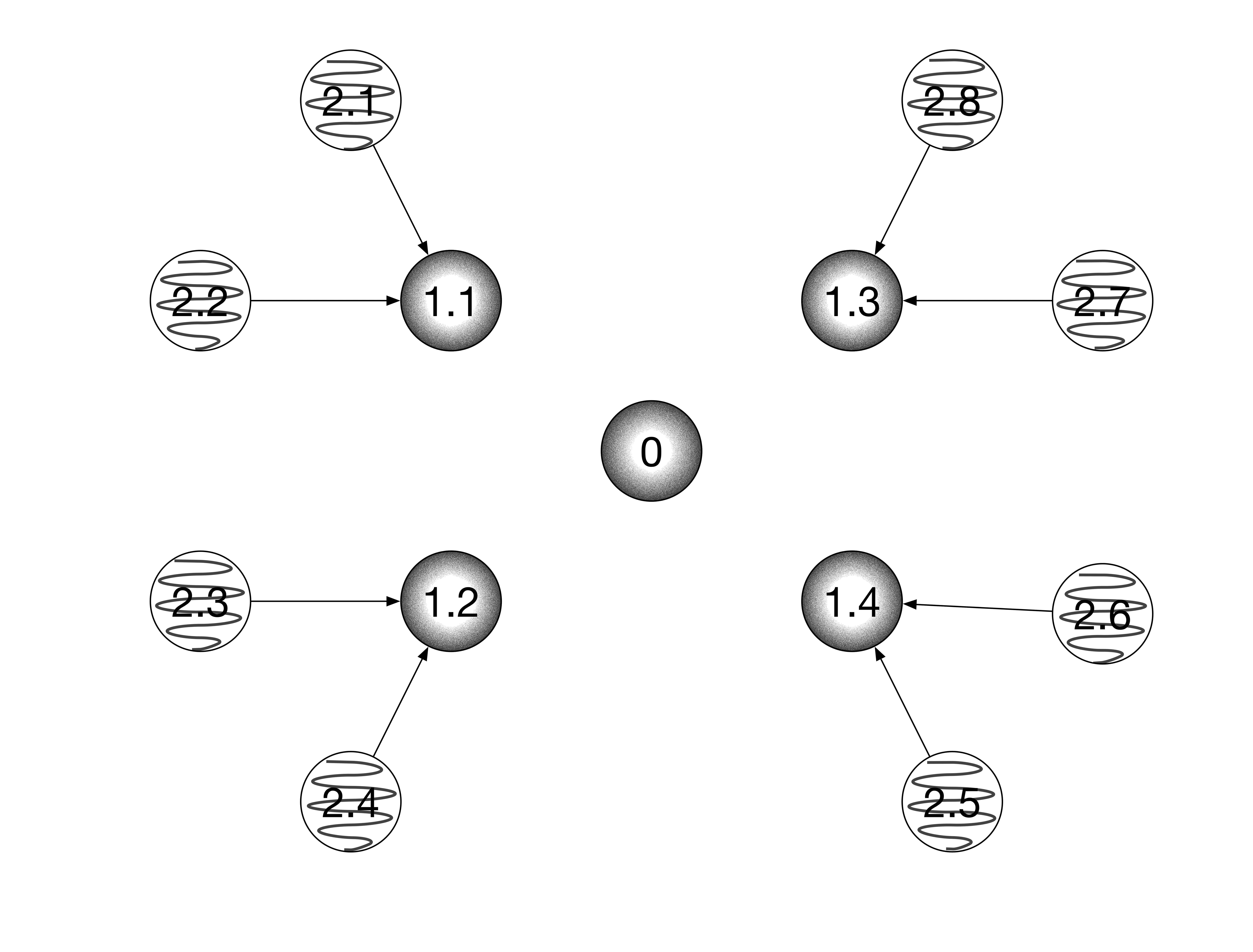}
    \caption{}
    \label{fig:examplesfig5}
  \end{subfigure}%
  \begin{subfigure}{.25\textwidth}
    \centering
    \includegraphics[width=.8\linewidth]{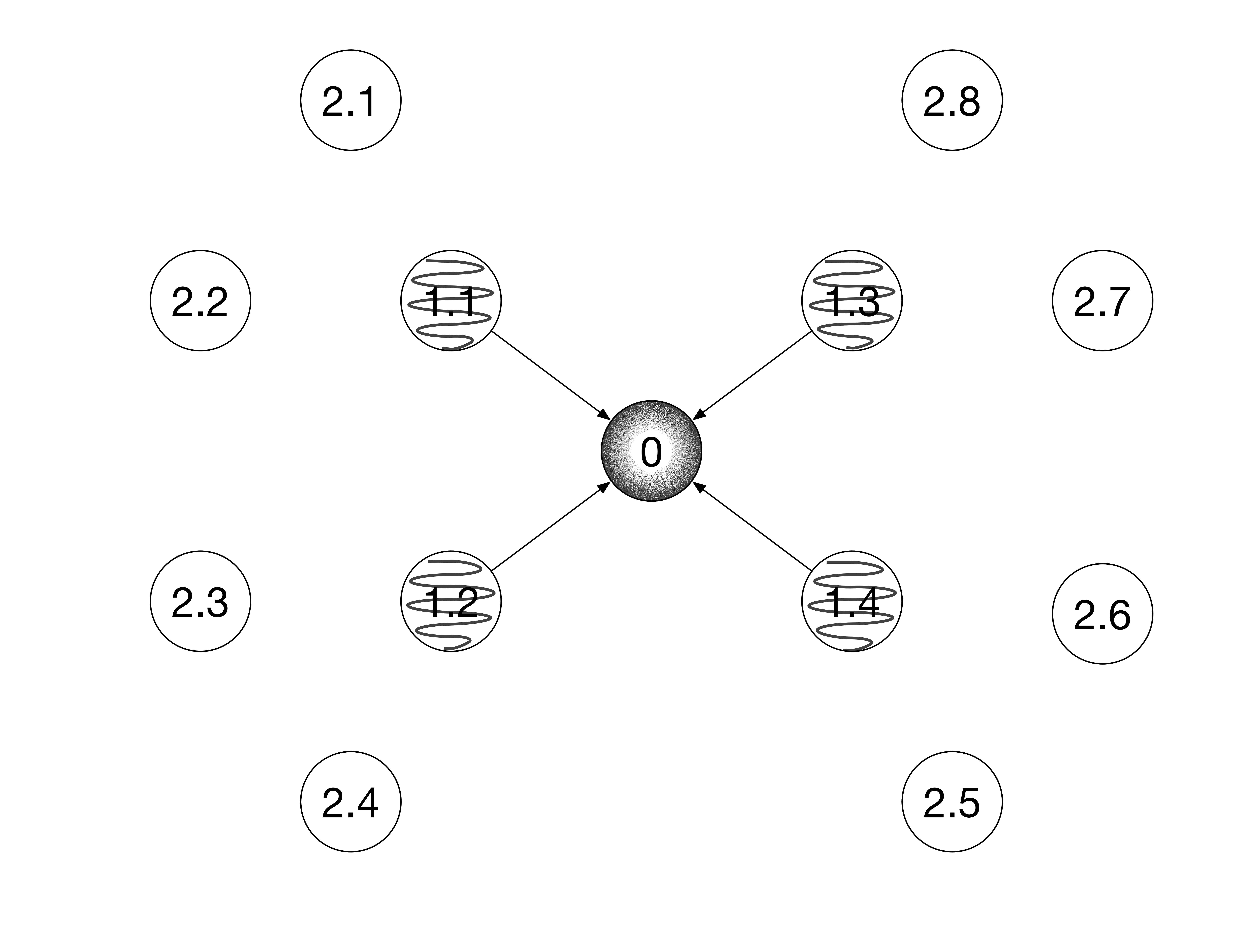}
    \caption{}
    \label{fig:examplesfig6}
  \end{subfigure}%
  \begin{subfigure}{.25\textwidth}
    \centering
    \includegraphics[width=.8\linewidth]{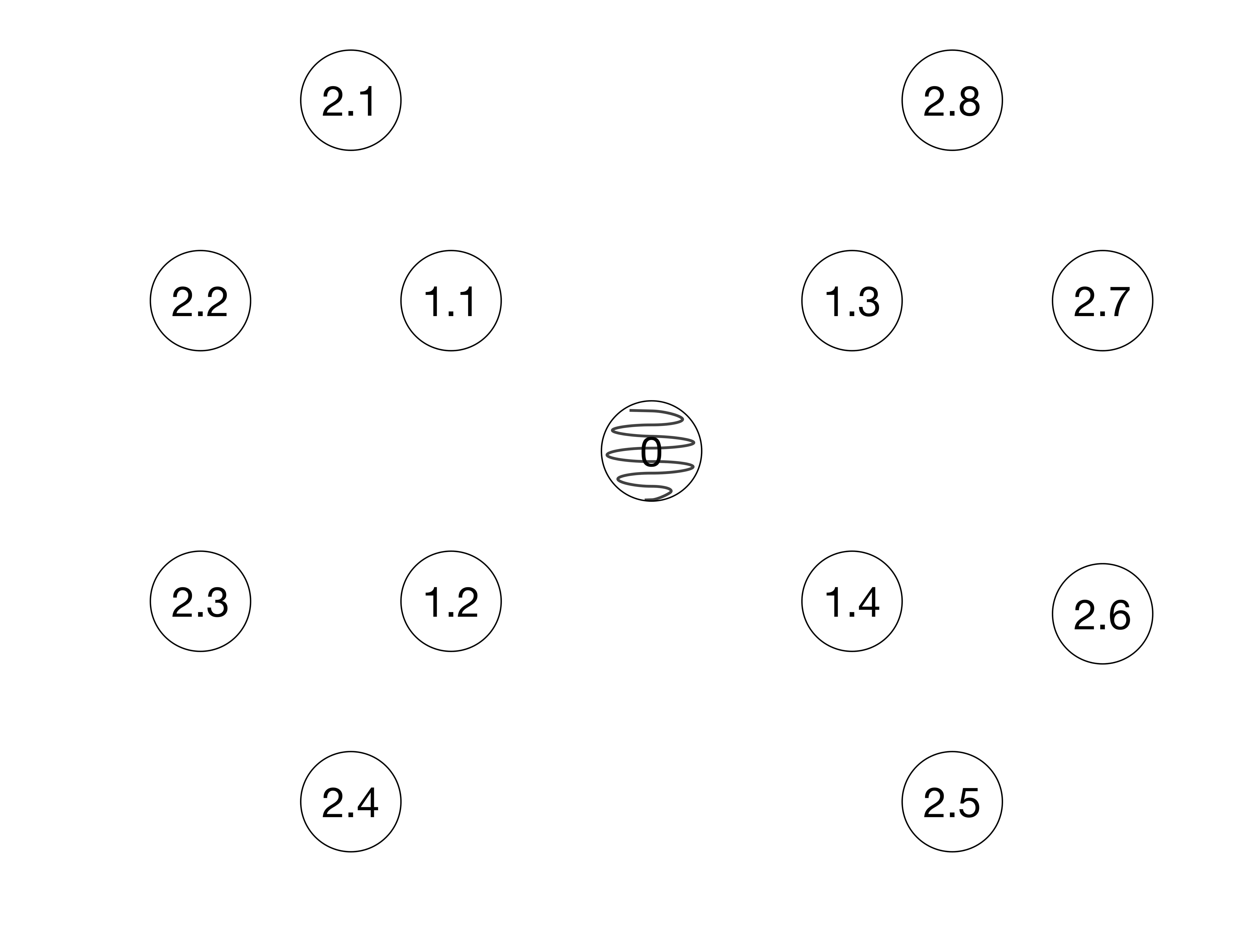}
    \caption{}
    \label{fig:examplesfig7}
  \end{subfigure}%
  \begin{subfigure}{.25\textwidth}
    \centering
    \includegraphics[width=.8\linewidth]{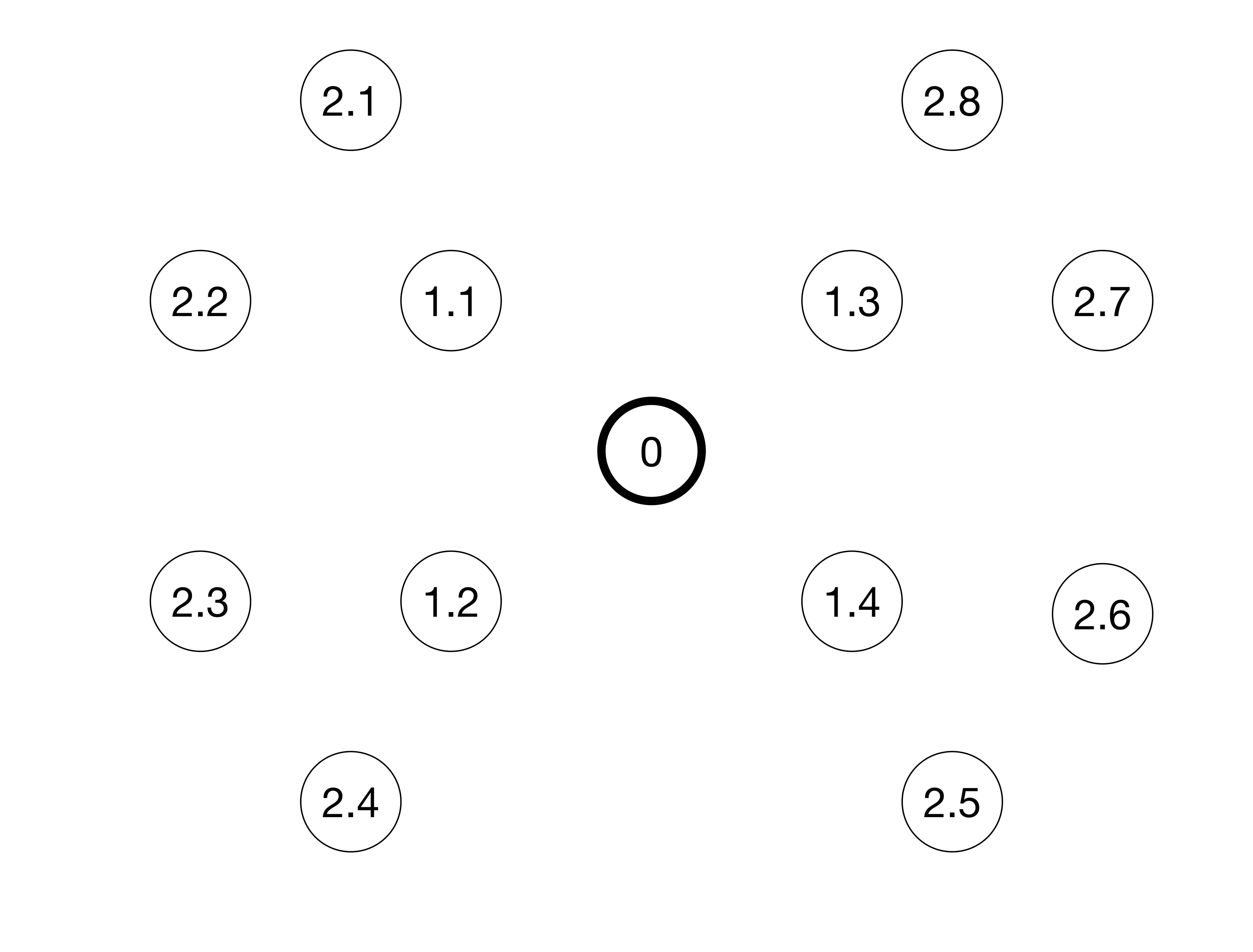}
    \caption{}
    \label{fig:examplesfig8}
  \end{subfigure}%
  \caption{Example of the hybrid monitoring approach}
  \label{fig:example}
\end{figure*}

The overall process is exemplified in Figure \ref{fig:example}.
In Fig. \ref{fig:examplesfig1}, the network is in the initial state.
Node 0 starts and changes its state to query state and it disseminates the query to neighbors within its range, as it can be seen in Fig. \ref{fig:examplesfig2}.
The query requests the computation of the value $f(x)$ and contains network information for the VHT.
Then each of these nodes repeats the same process, changes its state to query state and disseminates the query.
If we focus on one particular node, e.g., 1.1, we can see that the query is communicated to 2.1 and 2.2, as it can be seen in Fig. \ref{fig:examplesfig3}.
After that, all the network converges into the query state, as observed in Fig. \ref{fig:examplesfig4}.
Since the edge nodes are reached, these nodes change from query state to aggregate state and send their aggregation results to the parent node.
If we look at nodes 2.1 and 2.2 in Fig. \ref{fig:examplesfig5}, both share node 1.1 as the parent node, so each one of them sends their information to the parent node.
Then in Fig. \ref{fig:examplesfig6}, we can see that 2.1 and 2.2 are removed from the VHT, and that 1.1 changes his state to aggregate state.
In this moment, node 1.1 aggregates the results from 2.1, 2.2 and itself, and sends it to his parent node, i.e., node 0.
Then the process repeats itself with node 0 in Fig. \ref{fig:examplesfig7}.
Finally in Fig. \ref{fig:examplesfig8}, the network converged the monitoring process and node 0 has the global view of the monitored property.
The derived VHT that was built to calculate the queried value, is dissolved and no longer valid.

\subsection{Hybrid Architecture} \label{hybridArchitecture}

\subsubsection{Virtual Hierarchical Topology}

In our approach, a virtual hierarchical topology (VHT) for a time window is built to process a hierarchical based
aggregation of the monitored values through the network. The VHT concept has been introduced in \cite{huang2005virgo} and adapted by Google in one of his patents for cache nodes \cite{Goooglepatent}. 
The advantage of the VHT is its simple packet forwarding configuration. 
Each child node only forwards data packets to its parent node. 
The message will propagate in such a manner until it reaches the root node. 
The following steps summarize the VHT construction:\\
1- Each source node (chosen by the experts to aggregate the collected monitored values) sends a query to its neighbors. A timestamp information labels the time window.\\
2- Each node (that is not an edge node) receiving a query forwards it if not already received before. The hierarchy father/child and the timestamp are stored.\\
3- An edge node receiving a query does not forward it.


\subsubsection{Query State} \label{queryState}

The query state refers to the process of propagating in an epidemic way the monitoring packet.
This state goal is to disseminate the query and the VHT layout to allow the nodes in the network to do an accurate and efficient aggregation in the next state.
This query will be forwarded in an epidemic approach to the nodes in the relay set.
Lets call $A^1$ to the node sending the query and $A_i^1$ to the receiving nodes of the query.
In the exchange of information, $A^1$ will be the father node and $A_i^1$ will be the child nodes of $A^1$, since they will resemble the hierarchy of how the query is being propagated.
This will happen recursively until the edge of the network is reached.
This process is what creates the VHT.
It is a virtual representation, since given the mobility, it will not be able to keep the same topology physically.
The packet is explained in depth in Section \ref{results}, containing the query itself but also the information to generate the VHT.
This will communicate all the network information to create the VHT, which is the foundation of the following state of the network.
This process will go on until a node on the edge of the network is reached.

Along this state, there are some specific challenges to discuss.
\begin{enumerate*}[label=(\roman*)]
  \item The first challenge is if a node receives more than one monitoring packet once it is already in a monitoring state.
  This implies that a node is in the range of one or more probable parent nodes.
  For this, the node will take the first monitoring packet and will discard all the subsequent monitoring packets.
  \item The second challenge is, what if the propagation of the query is interrupted by a node that remains in a cyclic state. 
  For this, we introduce a timeout for the packet to avoid these problems. 
  This can be considered a time-to-live for the packet. 
  The idea is to provide a mechanism to avoid loops in the communications.
  This means that a node, because of a race condition, cannot continue disseminating the monitoring packet, since all the nodes in his relay set are child nodes of another node in the VHT.
  For this problem, the timeout will be triggered and once reached, this node will start the aggregation process by sending its result to the parent node.
  This way avoiding infinite loops in the VHT.
  \item The third challenge is the broadcast of the packet itself.
  Due to the nature of the simple epidemic dissemination approach, a packet will be forwarded to the next hop of nodes but also to the parent node (because of the medium).
  To make good use of this, we decided that this will work as an acknowledgment of the child node to the parent node.
  This way, the parent node will receive $n$ amount of acknowledgments and he will know how many packets he should wait for before changing to the aggregate state.
  We are assuming that the sender and the receiver nodes of this query packet will be within range to make a successful exchange of packets. After this is achieved, this nodes could get out of range and this will not affect our algorithm.
\end{enumerate*}

\subsubsection{Aggregate State} \label{aggregateState}

Once the data is disseminated up to the edge of the network, the edge nodes will change from query state to aggregate state and will start sending recursively their information to their parent up to the start node.
This process will be an aggregation of all the data of a node and his children in order to collect the monitored values.
The aggregation will be computed in a hierarchical manner with a combination when required of a gossip approach.
A node will compute based on his own observations the result of the function $f(x)$ that received from the query state.
This information will be aggregated with the same information of the child nodes.
In case of edge nodes, where this state starts, it will be done only with information from themselves.

Along this state, there are specific challenges to discuss.
\begin{enumerate*}[label=(\roman*)]
  \item The first one is when a parent node and a corresponding child node goes out of range from when they first met.
  When the child node sends an aggregate type of message and receives no acknowledgment it will trigger a route packet to the corresponding node.
  Depending on the node mobility it will be the time and the hops it will take in order to find the node.
  For this, we will rely on the gossip routing protocol of our approach.
  This will try to reach the parent node using a random approach for gossiping the packet.
  This means that the forwarder of the packet will be selected randomly, from the available neighbors at the moment.
  \item The second challenge is when a parent node is offline.
  It means that a child node already tried to route the package to him and still had no answer.
  For this, we propose that in the query state, a set of nodes are communicated to every child node for them to have an alternative path.
  Since the child node will have the relay set of the parent node, he will fall back into one of these nodes to send the information.
  This other node will continue the aggregating process.
  Since it is a hierarchical approach, the parent will send the information about his parents in the VHT.
  \item The third challenge is when a node receives a grandchild node aggregate information.
  For this, the node will assume that the child node is offline and that he will be aggregating that information.
  Given that the node does not know the information of how many grandchild will send information, he will also rely on the timeout before he sends his own aggregate information.
  For every grandchild packet he receives, he will restart the timeout to give time for additional packages.
  If the timeout is reached, he will continue with his aggregation process.
  If more information is received, 
  he will forward it to his parent and they will be aggregated ultimately by an upper node in the hierarchy 
  (eventually by the start node).
\end{enumerate*}


\section{Experiments} \label {results}

To support our architecture, we start by defining a protocol in a semi-formal way through a state 
automaton and also by defining the packet data sent over the network.
After this we implemented and conducted some tests to provide the support for the implementation.
In future works, we plan to enhance the tests to measure the scalability and the robustness under different node densities, node populations and mobility among other measures.

\subsection{Protocol Definition} \label{protocolAutomata}

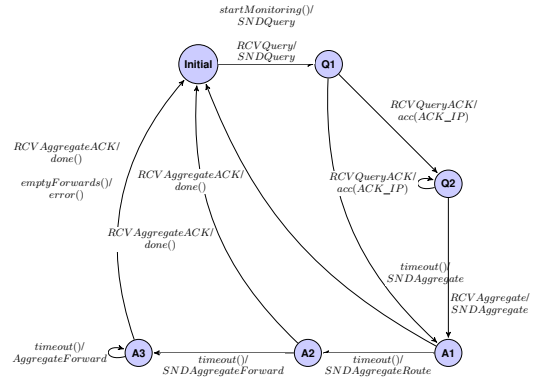
\begin{figure}
\centering

\scalebox{0.45}{%
\begin{tikzpicture}[->,>=stealth',shorten >=1pt,auto,node distance=5cm,
  thick,main node/.style={circle,fill=blue!20,draw,
  font=\sffamily\small\bfseries,minimum size=8mm}]

  \node[main node] (I) {Initial};
  \node[main node] (Q1) [right of=I,node distance=3.85cm] {Q1};
  \node[main node] (Q2) [below right of=Q1] {Q2};
  \node[main node] (A1) [below of=Q2] {A1};
  \node[main node] (A2) [left of=A1,node distance=4.15cm] {A2};
  \node[main node] (A3) [left of=A2] {A3};

  \path[every node/.style={font=\sffamily\small,
      fill=white,inner sep=1pt}]

  (I) edge [] node[align=center] {$startMonitoring()$/\\$SND Query$\\\\$RCV Query$/\\$SND Query$} (Q1)

  (Q1) edge [] node[align=center] {$RCV QueryACK$/\\$acc(ACK\_IP)$} (Q2)
  (Q1) edge [bend right=25] node[align=center] [near end] {$timeout()$/\\$SND Aggregate$} (A1)

  (Q2) edge [loop left] node[align=center] {$RCV QueryACK$/\\$acc(ACK\_IP)$} (Q2)
  (Q2) edge [] node[align=center] [near end] {$RCV Aggregate$/\\$SND Aggregate$} (A1)

  (A1) edge [] node[align=center] {$timeout()$/\\$SND AggregateRoute$} (A2)

  (A2) edge [] node[align=center] {$timeout()$/\\$SND AggregateForward$} (A3)

  (A3) edge [loop left] node[align=center] {$timeout()$/\\$Aggregate Forward$} (A3)

  (A3) edge [bend left=30] node[align=center] {$RCV AggregateACK$/\\$done()$\\\\$emptyForwards()$/\\$error()$} (I)
  (A2) edge [bend left=25] node[align=center] {$RCV AggregateACK$/\\$done()$} (I)
  (A1) edge [bend left=20] node[align=center] [near end] {$RCV AggregateACK$/\\$done()$} (I);

\end{tikzpicture}
}

\caption{State machine definition of our protocol} \label{fig:automata}
\end{figure}

The protocol definition, depicted in Figure \ref{fig:automata}, shows the expected behavior of the protocol to support as base ground for the hybrid monitoring architecture.
The set of states is $Q = (Initial, Q1, Q2, A1, A3, A3)$.
Where $Initial$ is the initial state.
The states $Q1$ and $Q2$ refer to the query states of the network.
And the states $A1$, $A2$ and $A3$ refer to the aggregate states of the network.
The internal operations of the automaton are startMonitoring(), acc(IP), timeout(), done() and error().
The startMonitoring() refers to the process of starting the monitoring.
The acc(IP) refers to the process of the node of accumulating the IP of the acknowledgment messages source.
For the node to accumulate this IP, it is required that the parent of the source node is the receiving node itself. 
If a node receives an acknowledgment message with a different parent of the source node, it means that other nodes are within range but the messages do not correspond to the node itself.
This is used to identify while the query is propagating if there are child nodes available for a given node.
If a node does not receive any acknowledgment, he will continue the monitoring process by using a timeout.
The timeout() refers to the process of counting time since the last package received.
The amount of time to wait will be transmitted through the protocol itself so it can be determined by the root node.
The done() refers to any internal process and manipulation of variables to restart the state of the node.
Meanwhile, error() refers to the process of not being able to send a message, which if it happens, it means that the node itself is out of the network range or a major outage is happening with the network.
So when this happens, the node is expected to log the problem, go back to the initial state and wait for further transmissions.
The input and output operations of the automaton are determined by sending (SND) and receiving (RCV) messages.
The possible messages to be sent or received are the query, query ack, aggregate, aggregate ack, agregate route and aggregate forward.
The query message refers to the query itself and the base ground of the query state.
For simplicity purposes, in the automaton, there is a distinction between the query and the query ack message.
But in reality, they are meant to be the same package but received by a different node. 
This is discussed in Section \ref{queryState}.
For example, if we refer to Figure \ref{fig:examplesfig3}, we can see that node 1.1 is sending the query message to nodes 2.1, 2.2 and 0.
For nodes 2.1 and 2.2, it means to receive a query message, but for node 0, given the properties transmitted in the package, then node 0 can compute that this message is the acknowledgment of his query message and that is why it is depicted as query ack.
The aggregate messages refer to the aggregation process and the same principle applies as the query messages.
The aggregate ack message is an aggregate message but received by a different node.
In this case, the receiver node of the aggregate ack is the child of a node, as opposite as the query ack which the receiver is the parent of a node.
Then we also have two extra messages which are the aggregate route and aggregate forward.
The aggregate route message refers to the process of routing a message through the network to the corresponding parent node in the VHT. 
As explained in Section \ref{aggregateState}, the idea is to make the hierarchical aggregation more robust through the addition of a gossip routing approach to route the package to the corresponding root node on the fly.
And finally, the aggregate forward message, whenever the parent node is not found, probably since the parent node went off line due to an outage or something similar.
In this case, the message will be forwarded to one of the nodes defined in the relay set, which will be populated by the grandparents and siblings.

\subsection{Packet Definition} \label{packetDefinition}

To have a successful communication, we need to define the monitoring packet.
The packet work equally in both states of the network, query and aggregate states, but different information are sent depending on the state containing a set of common properties. 
It needs to contain some basic information in order to be useful for the following nodes and hops.
The definition of such packet is done using json\cite{RFC7159}, given that it is one of the most used data interchange formats. 
For each state of the nodes, there is a set of transmitted properties.
These general properties are:

\begin{itemize}[leftmargin=0.4cm]
  \item Type: the type of message being sent, the set of values are query, aggregate, aggregate\_route, aggregate\_forward.
  \item Parent: the IP address of the parent node of the node sending the message.
  \item Source: the IP address of the node sending the message.
  \item Destination: the destination IP, this for most of the cases should be the parent IP, unless the parent node is off line, then it will be a relay set IP. 
  \item Gateway: the gateway IP, this is to support the routing which identifies the next hop of the packet.
  \item Timeout: the timeout in ms to avoid infinite loops and to support the aggregate state.
  \item Timestamp: a unique identifier calculated with the IP of the source concatenated with the Unix time of the system.
\end{itemize}

\noindent
For the query state, the specific properties transmitted are: 

\begin{itemize}[leftmargin=0.4cm]
  \item Function: the function $f$ to compute. We are considering the basic functions like CPU average usage, RAM average usage, or any other simple property.
  \item Relay Set: list of IPs for alternative paths. This list will contain at most three items. It should correspond to the parent, grandparent and great-grandparent node of a node. 
\end{itemize}


\noindent
For the aggregate state, the 
properties transmitted are:

\begin{itemize}[leftmargin=0.4cm]
  \item Result: the result of the aggregation of the function $f$. 
  This should be the aggregation of child nodes and the node itself.
  \item Observations: the number of observations aggregated. This value will be aggregated from the incoming aggregate packets.
\end{itemize}


\noindent
The json definition of the complete package is the following:

\begin{center}
{\tiny
\begin{lstlisting}[style=json]
{ "type": "query|aggregate|aggregate_route|aggregate_forward",
  "parent": "<parent IP>", "source": "<source IP>",
  "destination": "<destination IP>", "gateway": "<next hop IP>", 
  "timeout": <time in ms>, "timestamp": "<sourceIP + UnixTime>"
  "query": { "function": "<monitoring function>",
    "relaySet": [ "<IPs of the source rely set>" ] },
  "aggregate": { "outcome": "<monitoring result>",    
    "observations": <number of observations> } }
\end{lstlisting}
\vspace{-1em}
}
\end{center}

\subsection{Results} \label{implementation}

\begin{table}
  \caption{Scenario 1 parameters}\label{table:scenario1param}
  \vspace{-1em}
  \begin{center}
      \begin{tabular}{| l | l |}
      \hline
      \textbf{Parameter Name} & \textbf{Value} \\ \hline
      Number of nodes & 20, 25, 40, 50  \\ \hline
      Network Space & 400x400, 600x600  \\ \hline
      Network Positioning & Random  \\ \hline
      Emulation times & 100 (60s each with init time 40s)  \\ \hline
      Mobility & Random Waypoint Model at 5m/s \\ \hline
      \end{tabular}
  \end{center}
  \vspace{-2em}
\end{table}


We evaluate our proposal using an emulator built in-house based on DOCKEMU \cite{to2015dockemu}. 
This emulator (https://github.com/chepeftw/NS3DockerEmulator) is a combination between Docker and NS3.
Together they provide an environment highly scalable, replicable and robust to conduct experiments.
The testbed consisted in an implementation of the protocol in the language Go(https://github.com/chepeftw/Treesip).
The idea was to determine the convergence time, by which we mean the time it took from the moment that the monitoring started by the root node, to the moment that the root node was able to return a verdict.
Along this, we also looked at the number of observations collected, number of packets sent, size of the packets per run and the accuracy of the measurement.
The accuracy is defined as the ratio between the number of observations and the number of nodes.
We defined four scenarios with mobility.
We compare different number of nodes, space, speed, mobility pattern and routing protocol.
For all scenarios the MAC protocol is 802.11a, the application protocol is UDP and with a data rate of 54Mbps.
Each node had a range of $\approx$125m (theoretically by default in NS3 is 200m).
All scenarios were designed to test the convergence time and accuracy in a mobile environment. Given the mobility aspect of the scenarios, we assumed that it is difficult to achieve a fully accurate measurement. 
But the goal is to maximize the accuracy through the proposed architecture.

The emulator was running on top of an Amazon EC2 instance of type t2.medium and Ubuntu 16.04 LTS.
Versions in use were Docker 1.12.1, NS3.25 and Go 1.6.2.
The containers were running as a base Ubuntu 16.04 LTS and IPv4.
To collect the measurements we relied on the logs of the containers which were later parsed, exported to cvs and analyzed in R for further analysis.

\subsubsection{Scenario 1}

\imageref[0.40]{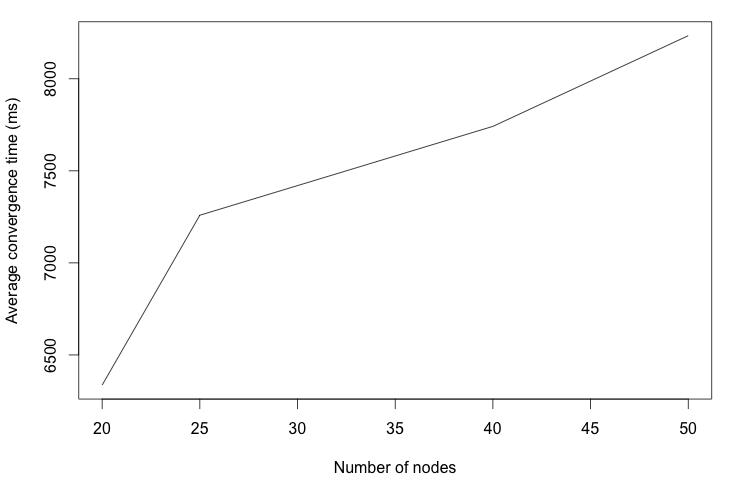}{Scenario 1 convergence results }{scenario1results}

As mentioned in Section \ref{methods}, the election process of the root node is out of the scope from this study, but for testing purposes, we decided to use different root node selected randomly to prove that it will work independently of who the root node is.
The parameters for this scenario can be seen in Table \ref{table:scenario1param}.
The results are summarized in Figure \ref{fig:scenario1results}.
We can point out that there is a clear relationship between the number of nodes and the time it takes to converge.
It is important to note that these times are subject to the defined timeouts for each run.
We based our timeouts in empirical data based on the number of nodes on different emulations.
Trying to find a safe value that will allow the network to converge without compromising the accuracy.
In overall, the average converge time it is approximately $\approx$7.51s.
About the number of packets sent, we empirically assumed that it would increase depending on the number of nodes given the gossip aggregation routing messages. 
For 20 nodes the average per run was 89 packets, for 25 nodes it was 138 packets, for 40 nodes it was 226 packets and for 50 nodes it was 280 packets.
The average message size for all runs is approximately $\approx$98 bytes.
The results collected by the monitoring process are approximately $\approx$0.8 accurate.

\subsubsection{Scenario 2}

For this scenario, we utilized 25 nodes, a 350x350m space and the random waypoint mobility model.
We tested the speeds 5m/s, 10m/s and 15m/s.
For each one, we ran the simulations 25 times.
The results are summarized in Table \ref{table:scenario2results}.
We assumed that the converge time might remain stable, but the accuracy could drop for a higher speed.
We can observe that the nodes converge about the same amount of time in the first two scenarios but for the third one increase significantly ($\approx$3.2s).
And contrary to our preliminary expectations, the accuracy remained approximately the same for 5m/s and 10m/s, and for 15m/s it slightly decreased.
The algorithm is capable of converging with nodes moving at different speeds with a stable accuracy.
This shows promising results for higher mobility scenarios.

\begin{table}
  \caption{Scenario 2 results}\label{table:scenario2results}
  \vspace{-1em}
  \begin{center}
      \begin{tabular}{| l | l | l | l |}
      \hline
       & \textbf{5m/s} & \textbf{10m/s} & \textbf{15m/s} \\ \hline
      Avg convergence time (ms) & 6336.12 & 6245.542 & 9419.9  \\ \hline
      Avg observations (\# nodes) & 20 & 19.8 & 18  \\ \hline
      Accuracy & $\approx$0.80 & $\approx$0.79 & $\approx$0.72  \\ \hline
      \end{tabular}
  \end{center}
  \vspace{-1em}
\end{table}

\subsubsection{Scenario 3}

For this scenario, we utilized 25 nodes, a speed of 5m/s and a 350x350m space.
We used the mobility patterns from NS3 of random waypoint model and random walk model.
For each mobility pattern we ran the simulations 25 times.
The results are summarized in Table \ref{table:scenario3results}.
We can observe that the convergence time has a difference of $\approx$2.5s, the average observations have a difference of 2 nodes and consequently the accuracy is different by 0.08.
This reflects, in our opinion, a fairly similar result in general.
But most importantly, it shows that it works in different mobility patterns whatever 
the properties are affected.

\begin{table}
  \caption{Scenario 3 results}\label{table:scenario3results}
  \vspace{-1em}
  \begin{center}
      \begin{tabular}{| l | l | l |}
      \hline
       & \textbf{Random Waypoint} & \textbf{Random Walk} \\ \hline
      Avg convergence time (ms) & 6336.12 & 8930.125  \\ \hline
      Avg observations (\# nodes) & 20 & 18  \\ \hline
      Accuracy & $\approx$0.80 & $\approx$0.72  \\ \hline
      \end{tabular}
  \end{center}
  \vspace{-1em}
\end{table}

\subsubsection{Scenario 4}

For this scenario, we utilized 25 nodes, a speed of 5m/s and a 350x350m space.
In this simulation, our intention is to compare the performance of our gossip routing approach and the protocol OLSR.
For each approach we ran the simulations 25 times, giving one minute for OLSR to have enough information.
The results are summarized in Table \ref{table:scenario4results}.
The convergence times are slower in our approach but more accurate, meanwhile in OLSR is faster but less accurate.
This results derives an interesting discussion.
If we use OLSR, our approach becomes dependent on the routing layer, but at the same time it delegates this responsibility and it allows our approach to focus on the monitoring process.
If we rely on our gossip routing approach, the solution becomes independent but it adds more complexity.
So there is a clear trade-off that should be considered.
After analyzing the logs and the scenarios we concluded that some scenarios were not so accurate because of the node reachability.
This is something we intend to study in depth in@ future works.

\begin{table}
  \caption{Scenario 4 results}\label{table:scenario4results}
  \vspace{-1em}
  \begin{center}
      \begin{tabular}{| l | l | l |}
      \hline
       & \textbf{Gossip} & \textbf{OLSR} \\ \hline
      Avg convergence time (ms) & 6336.12 & 4272.65  \\ \hline
      Avg observations (\# nodes) & 20 & 14  \\ \hline
      Accuracy & $\approx$0.8 & $\approx$0.56  \\ \hline
      \end{tabular}
  \end{center}
  \vspace{-2em}
\end{table}


\section{Related Works} \label{relatedworks}

MANET monitoring has been studied during several years for many objectives like their performances \cite{mehrotra2014performance}, to test them \cite{Maag:2008:ITM:1363686.1364148}, their security \cite{huang2003cooperative} and more recently their energetic efficiency \cite{palaniappan2015energy}. While there have been many works 
on decentralized monitoring approaches, 
There are multiple categories of decentralized monitoring, but
the two more prominent are gossip and hierarchical based \cite{stingl2012benchmarking}.
In the gossip based categorization, we can discuss about Gossipico \cite{van2012gossip}.
This is an algorithm to calculate the average, the sum or the count of node values in a large dynamic network.
The main focus is to count nodes in networks since this information can be useful for the performance of routing protocols, topology parameters or simply to determine how large a MANET is.
The combination of two mechanisms, count and beacon, provides the networks nodes counting in an efficient and quick way.
This algorithm was tested in static networks with dynamic scenarios.

After this study, there is a different paper that studied the same approach but totally focused on ad-hoc networks.
This algorithm, relying on Gossipico, is called Mobi-G \cite{stingl2014mobi}.
Mobi-G is designed for urban outdoor areas with a focus on pedestrian that moves around by foot.
The idea is to create the global view of an attribute, which is built ideally incorporating all the nodes in the network.
And this global view, disseminates it to all the nodes in the network to inform the current system state.
Mobi-G can provide accurate results even for fluctuating attributes. It also can reduce the communication cost.
It does not suffer from long range connectivity. Nevertheless the accuracy decreases for an increasing spatial network size.

On the hierarchical categorization, we can mention BlockTree \cite{stingl2013blocktree} that proposes a fully decentralized location-aware monitoring mechanism for MANETs.
The idea is to divide the network in proximity-based clusters, which are arranged hierarchical.
Each cluster or block will aggregate the data respecting a property.
The algorithm requires that all nodes from the same cluster or block are reachable within one hop.
This process will repeat among the hierarchy until a global view is reached, and then this value will be disseminated through the network.
Even though the good performance, the average power consumption increases directly proportional to the spatial network size or node density.

In the literature, we can also find an interesting paper \cite{stingl2012benchmarking} regarding a benchmark between the both categories.
This study states that the three main key points in architectural description for decentralized monitoring mechanisms are: 
\begin{enumerate*}[label=(\roman*)]
	\item Topology, construction and maintenance
	\item Data collection
	\item Result dissemination.
\end{enumerate*}
The authors define quality aspects as performance, cost, fairness, scalability, robustness and stability.
They also provide different workloads for the tests: baseline (idealized conditions), churn, massive join or crash, increasing number of attributes and increasing number of peers.
Under ideal conditions, the hierarchical approach outperforms the gossip approach.
In the presence of sudden topology changes, the gossip approach performs well and is able to continue working robustly. 
On the other side, the hierarchical approach collapses during these changes.
Nonetheless, this also depends on the hierarchical approach, but a tree-based approach will collapse.
With vertical scalability workload the hierarchical approach has smaller increases in traffic regarding the performance, while the gossip approach must handle considerable amounts of traffic.
Another result from this benchmark is the difficulty of comparing both categories.

\section{Future Work} \label{futurework}

We intend to study the selection of the root node.
It could be based on location, energy, computing power and other parameters.
It could also be an autonomous process, proactively or reactively, or a manual process that needs to be performed by an operator.

We will be looking for specific metrics under different scenarios in a scalable way.
In particular, we expect to assess the performance in terms of convergence of communication packets with a high node speed.


Lastly we intent to analyze and take into consideration more complex functions.
We are interested in monitoring the interoperability within a MANET.
For this we need to define an optimal solution to propagate a more complex function through our query mechanism.
This requires to analyze multiple interoperability approaches to provide a proposal for monitoring this process.
To monitor interoperability we guess that it is needed to analyze not all observation points, 
but more specifically a subset of observation points.

\section{Conclusions} \label{conclusions}

We have presented in this paper a hybrid algorithm for monitoring decentralized networks.
It consists on the combination of gossip-based and hierarchical-based algorithms.
Providing a robust and scalable solution, taking advantage of the best qualities of both approaches for wider scenarios.
The algorithm works on top of two major states, the query and the aggregate.
The gossip-based approach is applied to the query state, to disseminate the query in an efficient way.
Once the query is propagated, the network changes to the aggregate state.
Besides, with the help of a time-based hierarchical approach, the computation of a global property is achieved.
Based on our study, we concluded the recommendation of a monitoring protocol and we defined the following: 
\begin{enumerate*}[label=(\roman*)]
	\item an automaton describing the protocol as a mathematical support. Which provides an efficient way of representing a protocol and a mathematical test ground.
	\item The protocol definition using json\cite{RFC7159}.
	\item The importance and the versatility of a novel distributed monitoring architecture through the combination of a monitoring protocol and a hybrid algorithm.
\end{enumerate*}
We designed a scalable and configurable testbed using NS3 and Docker, based on DOCKEMU \cite{to2015dockemu}, that illustrates the effectiveness of our approach for different number of nodes, speeds and mobility patterns. 
Our approach 
and results are highly promising.

\bibliographystyle{plain}
\bibliography{ms.bib}


\end{document}